\documentclass[lettersize,journal]{IEEEtran}
\usepackage{amsmath,amsfonts}
\usepackage{algorithmic}
\usepackage{algorithm}
\usepackage{array}

\usepackage{textcomp}
\usepackage{stfloats}
\usepackage{url}
\usepackage{verbatim}
\usepackage{graphicx}
\usepackage{cite}
\usepackage{subfigure}
\usepackage{cleveref}
\usepackage[table,xcdraw]{xcolor}
\usepackage{balance}
\usepackage{flushend}

\begin{document}

\title{NeuroAiR: Deep Learning Framework for Airwriting Recognition from Scalp-recorded Neural Signals}
\author{Ayush Tripathi, Aryan Gupta, Prathosh A.P., Suriya Prakash Muthukrishnan, and Lalan Kumar
\thanks{This work was supported in part by Prime Minister’s Research Fellowship (PMRF), Ministry of Education (MoE), Government of India with grant number PLN08/PMRF (to Ayush Tripathi).}
\thanks{This work involved human subjects or animals in its research. Approval
of all ethical and experimental procedures and protocols was granted by
the Institute Ethics Committee, All India Institute of Medical Sciences, New Delhi, India, with reference number IEC-267/01.04.2022,RP-55/2022.}
\thanks{Ayush Tripathi is with the Department of Electrical Engineering, Indian Institute of Technology Delhi, New Delhi - 110016, India(e-mail: ayush.tripathi@ee.iitd.ac.in).}
\thanks{Aryan Gupta is with the Department of Electrical Engineering, Indian Institute of Technology Delhi, New Delhi - 110016, India(e-mail: aryaryary23@gmail.com).}
\thanks{Prathosh A.P. is with the Department of Electrical Communication Engineering, Indian Institute of Science, Bengaluru - 560012, India(e-mail: prathosh@iisc.ac.in).}
\thanks{Suriya Prakash Muthukrishnan is with the Department of Physiology, All India Institute of Medical Sciences, New Delhi - 110016, India(e-mail: dr.suriyaprakash@aiims.edu).}
\thanks{Lalan Kumar is with the Department of Electrical Engineering,
Bharti School of Telecommunication, and,
Yardi School of Artificial Intelligence, Indian Institute of Technology Delhi, New Delhi - 110016, India(e-mail: lkumar@ee.iitd.ac.in).}}

\markboth{Manuscript Under Review}%
{}


\maketitle

\begin{abstract}

Airwriting recognition is a task that involves identifying letters written in free space using finger movement. It is a special case of gesture recognition, where gestures correspond to letters in a specific language. Electroencephalography (EEG) is a non-invasive technique for recording brain activity and has been widely used in brain-computer interface applications. Leveraging EEG signals for airwriting recognition offers a promising alternative input method for Human-Computer Interaction. One key advantage of airwriting recognition is that users don't need to learn new gestures. By concatenating recognized letters, a wide range of words can be formed, making it applicable to a broader population. However, there has been limited research in the recognition of airwriting using EEG signals, which forms the core focus of this study. The NeuroAiR dataset comprising EEG signals recorded during writing English uppercase alphabets is first constructed. Various features are then explored in conjunction with different deep learning models to achieve accurate airwriting recognition. These features include processed EEG data, Independent Component Analysis components, source-domain-based scout time series, and spherical and head harmonic decomposition-based features. Furthermore, the impact of different EEG frequency bands on system performance is comprehensively investigated. The highest accuracy achieved in this study is $44.04\%$ using Independent Component Analysis components and the EEGNet classification model. The results highlight the potential of EEG-based airwriting recognition as a user-friendly modality for alternative input methods in Human-Computer Interaction applications. This research sets a strong baseline for future advancements and demonstrates the viability and utility of EEG-based airwriting recognition.

\end{abstract}


\section{Introduction}  

\subsection{Background}

Communication is a vital skill for humans, and with the rise of digital devices in recent years, our interaction with such devices has also increased significantly. To cater to this growing demand, new methods of input need to be developed. These alternative input methods should be wearable and integrate with human cognition seamlessly \cite{MIT, FB, Meta}. In particular, virtual reality devices that directly present information to the user's eyes require innovative input solutions, as external peripherals may not be feasible. While speech-based input is commonly used \cite{SpeechVR}, it suffers from challenges like noise, reverberations, and speech disorders that can degrade system performance \cite{Moore2018}. Additionally, speech-based input may not be suitable for maintaining privacy in public places. Gesture recognition emerges as an alternative solution that addresses privacy concerns and enables silent information transmission \cite{yasen2019systematic,cheok2019review}. However, it also has limitations, such as relying on a fixed dictionary of gestures, which restricts the range of interaction. Moreover, users need to learn and remember the specific set of gestures, adding a cognitive burden. Airwriting recognition represents a specialized form of gesture recognition that overcomes the aforementioned limitations. Airwriting involves the act of writing or drawing in the air and can be captured using different modalities, including inertial sensors \cite{tripathi2021sclair, tripathi2022imair}, electromyography \cite {tripathi2023surfmyoair, tripathi2023tripceair}, and computer vision techniques \cite {MUKHERJEE2019217, kim2021writing}. 

This study focuses on evaluating the effectiveness of a non-invasive brain-computer interface (BCI) for recognizing airwriting gestures. BCI offers a direct connection between the electrical activity of the brain and an external device, enabling enhanced communication and control capabilities. Driven by evolving algorithms and a desire to improve communication and interaction between humans and technology, the field of BCI has witnessed significant advancements in recent years \cite{wolpaw2012brain, lotte2018review}. BCI systems have the potential to empower individuals with disabilities, enabling seamless interaction with others. Language production systems based on brain signals have received considerable attention in BCI research. BCIs can be broadly classified into invasive and non-invasive methods based on the data acquisition approach \cite{buzsaki2012origin}. Invasive methods involve implanting sensors inside the brain to capture neural activities. However, these methods are costly, challenging to implement, and cause discomfort to the user. On the other hand, non-invasive methods record signals by placing sensors on the scalp \cite{islam2023recent}. Non-invasive methods are more affordable and user-friendly. Electroencephalography (EEG) is a commonly used non-invasive technique to measure brain electrical activity and has been extensively applied in BCI systems \cite{panachakel2021decoding,lazarou2018eeg}. Motivated by this, the current study investigates the recognition of airwriting gestures at the English alphabet level using EEG signals. This approach allows users to interact with the system without learning a new set of gestures. Furthermore, as the combination of letters can form a wide range of words, airwriting offers users extensive interaction capabilities. This user-friendly system has the potential to be easily adopted by a larger portion of the population, making it more accessible and versatile.

\subsection{Related Work}

The exploration of BCI systems aiming to emulate the language production system has been a subject of considerable interest in previous research \cite{zhang2018review}. One notable example is the BCI speller, where users observe flickering letters designed to evoke P300 or steady-state visual evoked potentials related to specific letters \cite{fazel2012p300, medina2020p300}. However, perception-based paradigms like these have encountered practical challenges, particularly in terms of participant fatigue induced by prolonged usage of such devices.

The imagery-based paradigm is widely recognized and utilized in the field of BCI, where neural information generated by internally imagined actions is translated into control signals. Various systems have been developed based on this paradigm, including those focused on specific actions like reaching, grasping, and hand movements \cite{alimardani2018brain, 10124364, 9837422}. This paradigm has also been extended to language-related tasks. For instance, Kumar et al. \cite{kumar2018envisioned} proposed an envisioned speech recognition framework using EEG signals, employing a random forest classifier and handcrafted features to classify digits, characters, and objects. Subsequently, a Deep Spatio-Temporal model was introduced to further enhance the system's performance \cite{kumar2021deep}. However, the efficacy of motor-imagery-based BCI systems can be hindered by the low signal-to-noise ratio of EEG signals \cite{rashid2020current}.

In contrast, motor-execution-based paradigms require participants to physically perform the intended movements, which are then decoded using EEG signals. This paradigm has been applied to various tasks such as arm movement \cite{buerkle2021eeg}, grasp-and-lift \cite{jain2022premovnet, luciw2014multi}, and biceps curl trajectory estimation\cite{saini2023bicurnet}. Pei et al. \cite{pei2021online} proposed a system to identify handwritten letters using EEG signals recorded from the scalp. Participants were instructed to repeatedly write the sentence 'HELLO, WORLD!' on a tablet while their EEG signals were captured using a $32$-channel device. Independent component analysis was employed to extract features from the EEG signals, which were then input into a convolutional neural network (CNN) model for letter recognition.

The use of direct cortical recordings through electrocorticography (ECoG) presents a distinct advantage compared to other signal acquisition methods. ECoG signals are less susceptible to artifacts and the spatial mixing of different source activities, allowing them to preserve the original neural activity reliably. Recent notable advancements have showcased the potential of ECoG signals in various language-related applications. Several studies have demonstrated that ECoG signals recorded during speech production or perception can be utilized to synthesize the original speech or directly translate it into text with impressive levels of accuracy \cite{wilson2020decoding, sun2020brain2char}. Additionally, the decoding of ECoG signals during handwriting imagery has shown promising results in typing applications, achieving high performance in converting imagined handwriting into text \cite{willett2021high}. These studies have revealed a clear correspondence between neural dynamics and the patterns of fine motor control activity, providing a solid foundation for BCI applications in the language domain.

Based on the previous literature highlighting the relationship between cortical activities and actual or imagined movements, it is reasonable to assume a similar association between scalp-recorded EEG signals and the muscle movements involved in writing. This assumption forms the basis of our current study of airwriting recognition using EEG signals. Unlike traditional writing, airwriting involves the generation of characters in free space using wrist or finger movements. The absence of physical support for the fingers, as well as the lack of visual and haptic feedback, pose additional challenges to this task. Previous attempts at airwriting recognition have utilized different modalities such as inertial sensors \cite{tripathi2022imair,tripathi2021sclair}, EMG \cite{tripathi2023surfmyoair, tripathi2023tripceair}, and computer vision-based approaches \cite{kim2021writing, MUKHERJEE2019217}. However, to the best of our knowledge, the use of surface EEG signals for airwriting recognition has not been explored thus far. Inspired by the success of deep learning-based approaches in translating neural signals into specific tasks performed by individuals, the feasibility of surface EEG-based airwriting recognition is explored in this study.

\subsection{Objectives and Contributions}

The airwriting task involves the brain processing the intended letter and transmitting neural signals to the forearm muscles to execute the necessary movements for writing the character. However, there has been limited attention given to identifying airwriting using neural signals recorded from the scalp, which is the main focus of this study. First, a dataset consisting of surface EEG signals is collected from $10$ participants during the airwriting task. To the best of the authors' knowledge, this is the first dataset specifically designed for EEG-based airwriting recognition. Subsequently, various feature extraction techniques, in combination with different deep learning models, are employed to assess the performance of the airwriting recognition system. The key contributions of this article are outlined below:

\begin{itemize}
    \item The NeuroAiR dataset comprising of EEG signals recorded from $10$ subjects during the task of writing uppercase English alphabets is created.  
    
    \item The performance of airwriting recognition from EEG signals is analyzed using processed EEG, ICA components, source-domain scout time series, and harmonic decomposition-based features.  
    
    \item The effects of different EEG frequency bands: Delta, Theta, Alpha, Beta, and Gamma, for the task of airwriting recognition is thoroughly explored.  

    \item Analysis regarding reducing the number of ICA components and its effect on the airwriting recognition accuracy is also performed.  
    
    \item For the harmonic decomposition-based features, the effect of decomposition order on airwriting recognition performance is studied.

    \item All the source codes and the collected data used in this paper will be made available for usage in the HCI community. 
    
\end{itemize}

The remainder of the paper is organized as follows: Section II details the EEG data collection setup (Section II.A), processing of EEG signals (Section II.B), source-domain feature extraction(Section II.C), spherical and head harmonic decomposition (Section II.D), and the deep learning models used for classification(Section II.E). Details regarding the performed experiments, results, and discussion are presented in Section III, while Section IV concludes the paper.

\begin{figure*}[ht!]
  \centering
  \centerline{\includegraphics[width=0.85\linewidth]{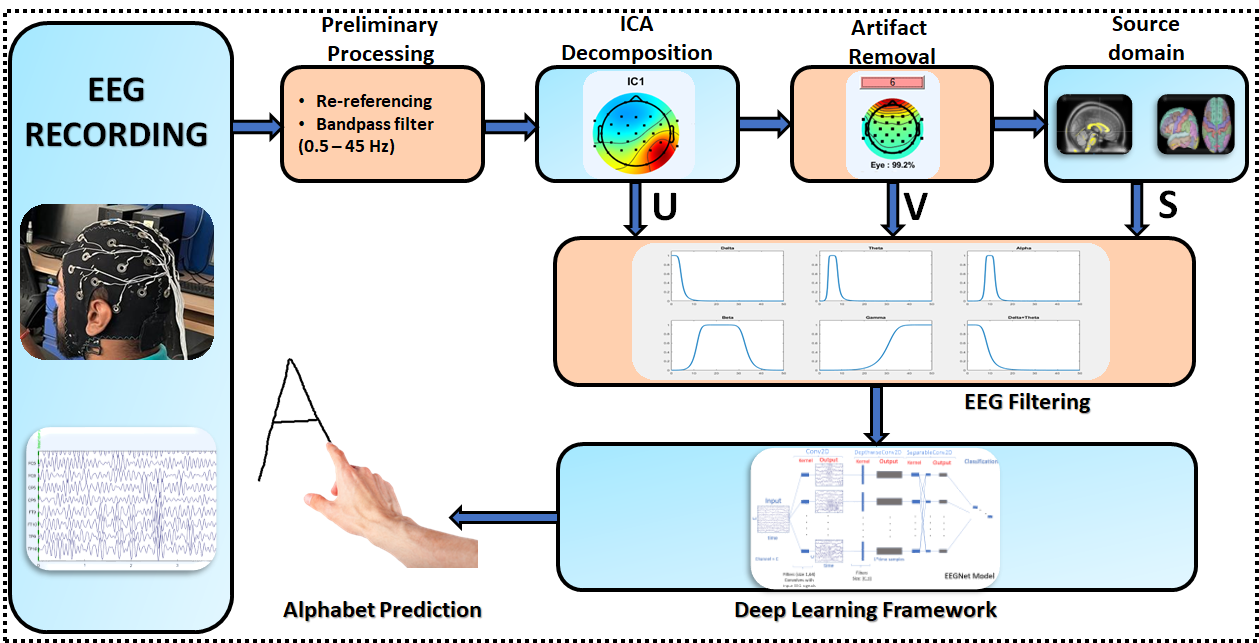}}
  \caption{Block diagram depicting an overview of the proposed EEG-based airwriting recognition system.} 
\label{fig:blockdiagram}
\end{figure*}

\section{Materials and Methods}

In this Section, the data recording procedure for EEG-based airwriting is presented along with the preprocessing steps, feature extraction, and the Deep Learning models used for classification. An overview of the steps involved in extracting the different sets of features utilized in the work is presented in Figure \ref{fig:blockdiagram}. 

\begin{figure}[!t]
    \centering{
    \includegraphics[width=0.75\linewidth]{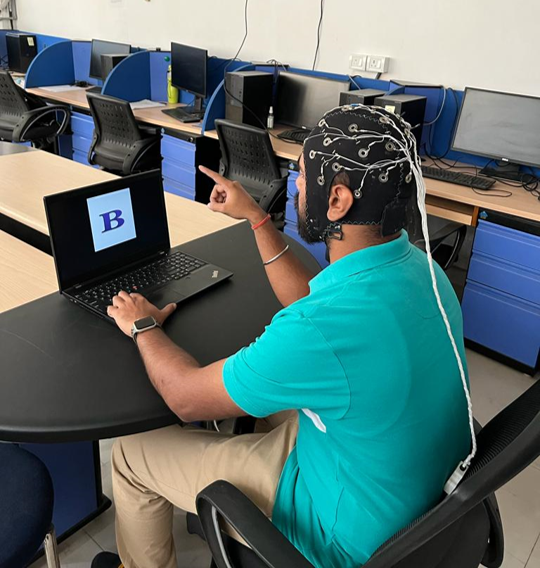}}
    \caption{Depiction of the EEG-based airwriting data collection setup.}
    \label{fig:datacollection}
\end{figure}

\subsection{EEG Data Recording}

EEG signals were recorded for the airwriting recognition task in the Multichannel Signal Processing Laboratory situated in the Electrical Engineering Department at Indian Institute of Technology Delhi. The data collection was carried out as per the Helsinki Declaration guidelines and the ethical approval was granted by the Institute Ethics Committee of All India Institute of Medical Sciences, New Delhi, India. $10$ healthy right-handed participants ($8$ male and $2$ female) with a mean age of $25.5 \pm 2.58$ years were recruited to take part in the experiment with their consent. All individuals included in the study had either normal vision or vision corrected to a normal level, and they confirmed that they had no previous history of mental disorders. Prior to the start of the recording, the entire experimental procedure was explained to the participant. The participants were given the chance to ask any questions, and the data recording session began only after all the queries of the participant were resolved. 

EEG signals were recorded with a 31-channel amplifier (LiveAmp, Brain Products GmbH, Germany) and referenced to the ground electrode. The recording was carried out using the Brain Vision Recorder software (Brain Products GmbH, Germany). EEG electrodes were placed on an electrode cap (EasyCap, Brain Products GmbH, Germany) as per the International 10-20 system. The signals were amplified and digitized at a sampling rate of 500 Hz using the amplifier. In order to ensure good signal quality, the impedance of each EEG electrode was maintained below $20 k\Omega$ at all times during the recording. 

The participants assumed a comfortable sitting position on a chair and positioned their right elbow firmly on the table. A user interface, created with the Tkinter module in Python, was utilized to provide the participant with a visual prompt indicating the character to be written and to record annotations throughout the experiment. The participants were instructed to operate the interface using their left hand. Upon pressing the spacebar, a character was displayed on the screen, prompting the participant to write the corresponding character in the air using their index finger. In this manner, each participant wrote $100$ sets of the $26$ English uppercase alphabets. This resulted in a total of $26\times100 = 2600$ samples for every subject. Every letter of the alphabet was recorded separately, and after every five sets of recording, the participant was given a rest period. The duration of this rest period was not predetermined, allowing the participant to take as much rest as needed. This approach aimed to introduce greater variation between repetitions of the same letter. Additionally, during the rest period, the impedance of each EEG electrode was checked to ensure that it is less than $20 k\Omega$. The complete setup for data collection, including the EEG recording arrangement, is depicted in Figure \ref{fig:datacollection}.

\subsection{EEG Processing}

The EEG data were preprocessed using MATLAB \cite{MATLAB} and the EEGLAB plugin \cite{EEGLab}. First, the recorded data corresponding to all the $100$ sets for a subject were merged. Subsequently, several processing steps were applied to the data to make it suitable for feeding it as input to deep learning models for the task of airwriting recognition. The data processing pipeline comprised of (a) preliminary preprocessing, (b) Independent Component Analysis (ICA) decomposition of the EEG signals, (c) Artifact removal from the EEG signals, and (d) filtering into different frequency bands as presented in Figure \ref{fig:blockdiagram}.

\subsubsection{Preliminary preprocessing}

The first step involved in EEG data processing is to re-reference the signals to a common average reference. The main concept behind the average reference technique is to calculate the average signal across all EEG electrodes and then subtract this average from the EEG signal at each electrode for every time point. The underlying rationale for this approach is based on approximating the shape of the head as a sphere. By assuming that the total sum of potentials recorded from all sources inside the spherical head is zero, using the average reference provides a reference point that is electrically neutral. Subsequently, the re-referenced EEG signals are bandpass filtered using a zero phase non-causal Finite Impulse Response (FIR) filter. The passband edges of the filter are set to be $[0.5,45]$ Hz, while the transition bandwidth is taken to be $0.5$ Hz. This filter ensures that the DC drift component is filtered out from the EEG recordings. Additionally, the electrical interference is also filtered out from the recorded EEG signals. 

\begin{figure*}[ht!]
  \centering
  \centerline{\includegraphics[width=0.7\linewidth]{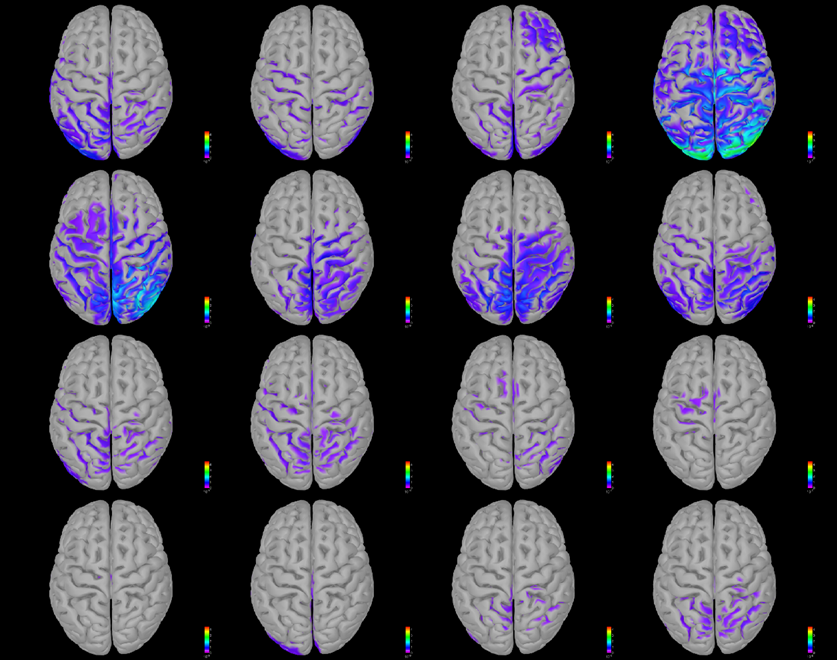}}
  \caption{Time sheet depicting the EEG source localization plots for equally spaced time intervals between $0$ to $2$ seconds. The Brainstorm toolbox \cite{Brainstorm} is used for generating the plots.} 
\label{fig:sLORETA}
\end{figure*}

\subsubsection{ICA decomposition}

After the preliminary processing of the recorded EEG signals, Independent Component Analysis (ICA) is applied to separate the EEG data into independent source activities. ICA is a specific algorithm for Blind Source Separation and has been successfully used to decompose EEG data in literature \cite{ICA}. The underlying assumption for applying ICA is that the EEG signals recorded by the electrodes are linear combinations of the underlying source activities. Mathematically, this can be represented as:

\begin{equation}
    [X] = [A][U]
\end{equation}
where $[X]$ denotes the 2D matrix corresponding to the EEG signals, $[U]$ denotes the ICA components, and $[A]$ is referred to as the mixing matrix. The primary goal of the ICA algorithm is to discover an unmixing matrix $[W]$ that allows for the extraction of statistically independent signals through the equation $[U] = [W][X]$. Mutual information serves as the criterion for measuring the statistical dependence of the signals. The ICA algorithm is implemented using the EEGLAB plugin, which generates the output in the form of independent components sorted by descending order of variance. It is to be noted that the components ($[U]$) obtained by ICA decomposition can be either neural or non-neural artifacts. The independent components are subsequently retained for artifact removal as well as for using it as a feature for identifying the airwritten letter.   

\subsubsection{Artifact removal}

A number of ICA components correspond to artifacts caused by muscle and eye movement. To preserve only the neural components and remove the artifacts, a two-step process is employed. Firstly, the artifact-related components are identified by classifying the components using the ICLabel function in EEGLAB. Components labeled as muscle or eye movement with a certainty exceeding $80\%$ are identified as artifact-related. To obtain clean EEG data, denoted as $[V]$, a modified version of the mixing matrix ($[A]$), referred to as $[\hat{A}]$, is used. In $[\hat{A}]$, the columns corresponding to the artifact-related components are set to zero. The clean EEG data $[V]$ is then computed as $[V] = [\hat{A}][U]$. Subsequently, this artifact-free EEG data is utilized as input for deep learning models for airwriting recognition.

\subsubsection{EEG Filtering}

EEG signals are generally subdivided into different bandwidths for analysis. In this study, the different EEG frequency bands and their effect during the airwriting task is analyzed. The clean EEG data is filtered into different components with specifications:

\begin{itemize}
    \item Delta: lowpass filter with passband edge at $4Hz$ and transition bandwidth of $2Hz$.
    \item Theta: badpass filter with passband edges at $[4,8]Hz$ and transition bandwidth of $2Hz$.
    \item Alpha: badpass filter with passband edges at $[8,12]Hz$ and transition bandwidth of $2Hz$.
    \item Beta: badpass filter with passband edges at $[12,32]Hz$ and transition bandwidth of $3Hz$.
    \item Gamma: highpass filter with passband edge at $32Hz$ and transition bandwidth of $8Hz$.
    \item Delta+Theta: lowpass filter with passband edge at $8Hz$ and transition bandwidth of $2Hz$.
\end{itemize}

In order to filter the EEG signals to the aforementioned frequency bands, zero-phase, non-causal FIR filters with desired specifications are used. The ICA components $[U]$ and clean EEG signals $[V]$ are filtered and stored for subsequent classification. These filtered EEG signals are also utilized for extracting source-domain and Spherical Harmonic Decomposition based features.

\subsection{Source-domain Features}

The fundamental concept of EEG source localization is to estimate primary cortical current densities based on the EEG data \cite{zaitcev2017eeg}. This approach addresses the issue of cross-electrode correlation caused by volume conduction effects. The pipeline for EEG source localization generally consists of two main stages: the forward problem and the inverse problem. A forward head model is used to establish the relationship between the voltages recorded on the scalp surface and the currents inside the head. Subsequently, the inverse problem is solved to determine the distribution of current sources that best fit the scalp potential while adhering to a set of predefined constraints. It should be noted that the number of EEG channels is typically around 30, while the number of current dipoles to be estimated is around 5000. This large disparity in the number of sensors and sources makes the EEG source localization severely underdetermined. The two stages involved in EEG source localization are further outlined in detail below.

\subsubsection{Forward Problem}

In the forward modeling stage, the lead field matrix is computed that describes the propagation of cortical source currents to the scalp electrode through different layers of conductivity. This involves utilizing the physics of EEG generation in conjunction with Neumann and Dirichlet boundary conditions to establish the relation between voltage recorded at the scalp and the cortical current densities.  Mathematically, this relationship can be expressed as:

\begin{equation}
    [V] = [L][\tilde{S}] + [Z]
\end{equation}
Here, $[V]$ is the processed EEG data, $[L]$ is the lead field matrix, $[\tilde{S}]$ is the cortical source current, and $[Z]$ is the noise perturbation matrix. 

In order to compute the head model, the numerical Boundary Element Method \cite{BEM} is utilized. More specifically, Brainstorm toolbox \cite{Brainstorm} is used for EEG source localization. First, the default  $cortex\_15002$, which consists of $15002$ vertices, is downsampled to $3000$ vertices using MATLAB's reducepatch resampling method. The ICBM152 MRI template is employed as the anatomy, and the head model is computed using OpenMEEG \cite{openMEEG} with parameters $\sigma_{scalp} = 1$, $\sigma_{skull} = 0.0125$, and $\sigma_{brain} = 1$.

\subsubsection{Inverse Problem}

In the inverse problem of EEG source localization, the lead field matrix $[L]$ is known as a prerequisite (from the head model computation). The aim then is to estimate the cortical source current signal $[\tilde{S}]$. The noise perturbation matrix $[Z]$ is added to the model to represent the error introduced by the solution of the inverse problem. In the current study, standard low-resolution electrical tomography (sLORETA) \cite{sLORETA} is utilized to obtain the solution to the inverse problem. sLORETA operates with the underlying assumption that neighboring voxels exhibit similar neural activity. This assumption is supported by the principle that nearby brain regions are more likely to contribute to the same neural process. By enforcing spatial smoothness, sLORETA aims to enhance the accuracy of source localization by promoting coherence and consistency in the estimated electrical activity across adjacent voxels. 

A sample source localization plot depicting the activation in the brain while writing a character is illustrated in Figure \ref{fig:sLORETA}. The plots are generated at equally spaced time stamps between $0$ to $2$ seconds. The entire brain is further grouped into $62$ regions based on the Mindboggle atlas \cite{klein2005mindboggle}. From the obtained cortical source current signals $[\tilde{S}]$, the signals belonging to a particular region are averaged out to obtain a time series corresponding to each of the $62$ regions. This yields the scout time series matrix $[S]$, which is then retained for utilization as input to deep learning models for airwriting recognition.

\begin{figure}[!t]
    \centering{
    \includegraphics[width=0.85\linewidth]{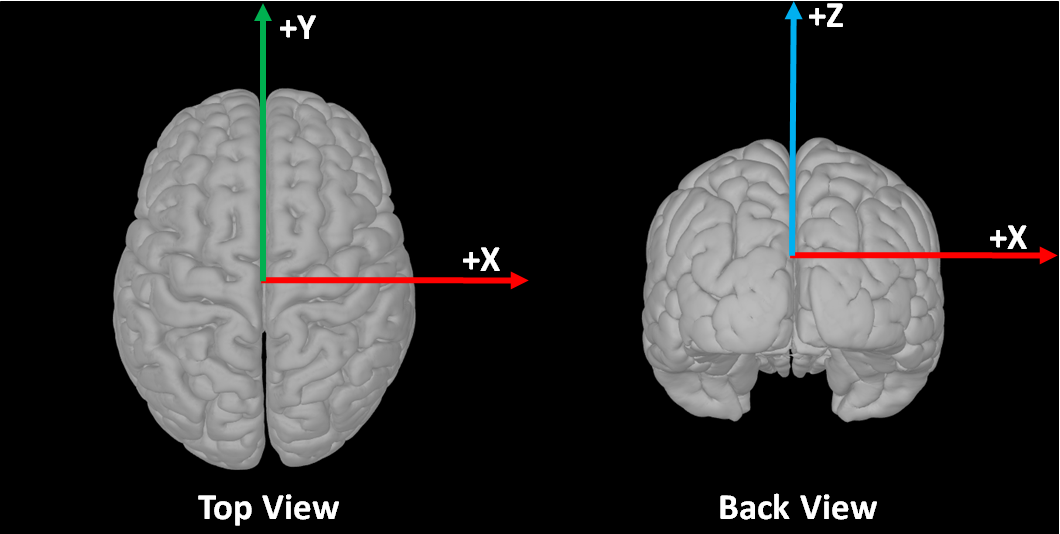}}
    \caption{Depiction of the coordinate axis convention used for Spherical and Head Harmonic Decomposition in the current study. The image template is taken from the Brainstorm toolbox \cite{Brainstorm}.}
    \label{fig:coordinatesystem}
\end{figure}

\subsection{Spherical Harmonic Decomposition based Features}

Spherical harmonics are a set of orthonormal basis for functions defined on the surface of a sphere \cite{jarrett2017theory}. Since the EEG signal can be viewed as a potential field on a surface that is almost spherical, it becomes feasible to perform Spherical Harmonic Decomposition (SHD) on the processed EEG signal \cite{wingeier2001spherical, TIM_BSL,giri2022anatomical}. 

At a particular time instant $t$, let the EEG potential at a location $\Omega = (R,\theta,\phi)$ originating from a source located at $\Omega_p = (r_p,\theta_p,\phi_p)$ be donated by $V(\Omega,\Omega_p,t)$. In the discussion, $R$ denotes the radius of the head, $\theta$, and $\phi$ are elevation and azimuth angles measured as per the coordinate system depicted in Figure \ref{fig:coordinatesystem}. The Spherical Harmonic Decomposition of the EEG potential is obtained as:

\begin{equation}
    V^{SH}_{nm}(\Omega_p,t) = \int_{\Omega}V(\Omega,\Omega_p,t)[Y_n^m(\Omega)]d\Omega
\end{equation}
In the aforementioned equation, $Y_n^m(\Omega)$ denotes the set of spherical harmonics of order $n$ and degree $m$. Assuming a finite decomposition order $N$, the order of spherical harmonics takes values $n \in [0,N]$, while degree takes values $m \in [-n,n]$. This results in a total of $(N+1)^2$ distinct harmonics. The maximum limit on the order is governed by the equation $N\leq(\sqrt{I}-1)$, where $I$ is the number of EEG sensors. Mathematically, the spherical harmonics are defined as\cite{TIM_BSL}:

\begin{equation}
Y_n^m(\theta,\phi) = \begin{cases}

(-1)^m\sqrt{2}K_n^m sin(|m|\phi)P_n^{|m|}(cos\theta) &\text{$m<0$}\\

(-1)^m\sqrt{2}K_n^m cos(|m|\phi)P_n^{|m|}(cos\theta) &\text{$m>0$}\\

K_n^0 P_n^0(cos\theta) &\text{$m=0$}\\

\end{cases}
\end{equation}
Here, $P_n^{m}(cos\theta)$ is the Associated Legendre Polynomial (ALP), and $K_n^m$ is a normalization constant which is computed as:

\begin{equation}
    K_n^m = \sqrt{\frac{(2n+1)(n-|m|)!}{4\pi(n+|m|)!}}
\end{equation}
In the discrete domain, Spherical Harmonic Decomposition in equation (3) can be rewritten as:

\begin{equation}
    V^{SH}_{nm}(\Omega_p,t) = \sum_{i=1}^{I} \gamma_iV(\Omega_i,\Omega_p,t)[Y_n^m(\Omega_i)]
\end{equation}
which can be equivalently written in matrix form as:

\begin{equation}
    [V^{SH}_{nm}] = [Y]^T\Gamma[V]
\end{equation}

Here, $\Gamma$ denotes the sampling weight matrix, which is taken to be the identity matrix in the current work. The SHD of the processed EEG signal, i.e., $[V^{SH}_{nm}]$, is extracted and fed to the deep learning models for the task of airwriting recognition. It is to be noted that the dimension of the EEG signal matrix $[V]$ is $I\times T$, while it is $(N+1)^2\times T$ for the SHD feature matrix. Here, $T$ denotes the number of timestamps which depends on the sampling rate and duration of the EEG signal considered.    

\textbf{Head Harmonic Decomposition:} The EEG acquisition sensors are positioned on the head, which takes on a shape that falls between a hemisphere and a sphere. It is, therefore, more suitable to establish a new set of basis functions specifically designed for the shape of the head instead of relying on the conventional Spherical Harmonics. Consequently, the concept of Head Harmonic was proposed in \cite{TIM_BSL} in order to meet this requirement. The head harmonics are defined as:

\begin{equation}
H_n^m(\theta,\phi) = \begin{cases}

(-1)^m\sqrt{2}\Tilde{K}_n^m sin(|m|\phi)\Tilde{P}_n^{|m|}(cos\theta) &\text{$m<0$}\\

(-1)^m\sqrt{2}\Tilde{K}_n^m cos(|m|\phi)\Tilde{P}_n^{|m|}(cos\theta) &\text{$m>0$}\\

\Tilde{K}_n^0 \Tilde{P}_n^0(cos\theta) &\text{$m=0$}\\

\end{cases}
\end{equation}
where, $\Tilde{P}_n^{m}(cos\theta) = P_n^{m}(1.33cos\theta-0.33)$ are shifted ALPs, while the modified normalization constant $\Tilde{K}_n^m$ is given by: 

\begin{equation}
    \Tilde{K}_n^m = \sqrt{\frac{(2n+1)(n-|m|)!}{3\pi(n+|m|)!}}
\end{equation}

Similar to SHD, the Head Harmonic Decomposition (HHD) in matrix form can be written as: 

\begin{equation}
    [V^{H^2}_{nm}] = [H]^T\Gamma[V]
\end{equation}
The HHD of the processed EEG signal i.e. $[V^{H^2}_{nm}]$, is used as input to the deep learning models for classification. Similar to SHD, the dimension of the HHD feature matrix is also $(N+1)^2\times T$.

Since, $(N+1)^2 \leq I$, therefore the SHD and HHD features offer the advantage of dimensionality reduction compared to the case where processed EEG signals are taken as input to the deep learning models.

\subsection{Deep Learning Frameworks}

Convolutional Neural Networks (CNN) based Deep Learning architectures have been successfully used to achieve high performance for several EEG-based BCI applications. Inspired by this, in the current study, three popular architectures, namely: EEGNet \cite{EEGNet}, DeepConvNet \cite{DeepConvnet}, and ShallowConvNet \cite{DeepConvnet} have been utilized. The model architectures are summarized in \cref{tab:EEGNet,tab:DeepConvNet,tab:ShallowConvNet}, while a detailed description is provided in the following subsections:

\begin{table}[t!]
\caption{Details of the EEGNet model architecture.}
\label{tab:EEGNet}
\centering
\scalebox{0.8}{
\begin{tabular}{cccc}
\hline
\textbf{Layer}   & \textbf{Kernel Size} & \textbf{\# of filters} & \textbf{Layer Parameters}            \\\hline\hline
Conv2D           & (1, 64)              & 8                      & Stride = (1, 1), Activation = Linear \\
BatchNorm        & -                    & -                      & -                                    \\
DepthwiseConv2D  & (31,1)               & -                      & Depth multiplier = 2                 \\
BatchNorm        & -                    & -                      & -                                    \\
Activation       & -                    & -                      & Activation = ELU                     \\
AveragePooling2D & (1, 4)               & -                      & Stride = (1, 4)                      \\
Dropout          & -                    & -                      & Rate = 0.5                           \\
SeparableConv2D  & (1, 16)              & 16                     & Zero Padding                         \\
BatchNorm        & -                    & -                      & -                                    \\
Activation       & -                    & -                      & Activation = ELU                     \\
AveragePooling2D & (1, 8)               & -                      & Stride = (1, 8)                      \\
Dropout          & -                    & -                      & Rate = 0.5                           \\
Flatten          & -                    & -                      & -                                    \\
Dense            & -                    & -                      & Neurons = 26, Activation = Softmax  \\\hline
\end{tabular}
}
\end{table}

\begin{table}[t!]
\caption{Details of the DeepConvNet model architecture.}
\label{tab:DeepConvNet}
\centering
\scalebox{0.8}{
\begin{tabular}{cccc}
\hline
\textbf{Layer} & \textbf{Kernel Size} & \textbf{\# of filters} & \textbf{Layer Parameters}            \\\hline\hline
Conv2D         & (1, 5)               & 25                     & Stride = (1, 1), Activation = Linear \\
Conv2D         & (31,1)               & 25                     & Stride = (1, 1), Activation = Linear \\
BatchNorm      & -                    & -                      & -                                    \\
Activation     & -                    & -                      & Activation = ELU                     \\
MaxPooling2D   & (1, 2)               & -                      & Stride = (1, 2)                      \\
Dropout        & -                    & -                      & Rate = 0.5                           \\
Conv2D         & (1,5)                & 50                     & Stride = (1, 1), Activation = Linear \\
BatchNorm      & -                    & -                      & -                                    \\
Activation     & -                    & -                      & Activation = ELU                     \\
MaxPooling2D   & (1, 2)               & -                      & Stride = (1, 2)                      \\
Dropout        & -                    & -                      & Rate = 0.5                           \\
Conv2D         & (1,5)                & 100                    & Stride = (1, 1), Activation = Linear \\
BatchNorm      & -                    & -                      & -                                    \\
Activation     & -                    & -                      & Activation = ELU                     \\
MaxPooling2D   & (1, 2)               & -                      & Stride = (1, 2)                      \\
Dropout        & -                    & -                      & Rate = 0.5                           \\
Conv2D         & (1,5)                & 200                    & Stride = (1, 1), Activation = Linear \\
BatchNorm      & -                    & -                      & -                                    \\
Activation     & -                    & -                      & Activation = ELU                     \\
MaxPooling2D   & (1, 2)               & -                      & Stride = (1, 2)                      \\
Dropout        & -                    & -                      & Rate = 0.5                           \\
Flatten        & -                    & -                      & -                                    \\
Dense          & -                    & -                      & Neurons = 26, Activation = Softmax  \\\hline
\end{tabular}
}
\end{table}

\begin{table}[t!]
\caption{Details of the ShallowConvNet model architecture.}
\label{tab:ShallowConvNet}
\centering
\scalebox{0.8}{
\begin{tabular}{cccc}
\hline
\textbf{Layer}   & \textbf{Kernel Size} & \textbf{\# of filters} & \textbf{Layer Parameters}            \\\hline\hline
Conv2D           & (1, 13)              & 40                     & Stride = (1, 1), Activation = Linear \\
Conv2D           & (31,1)               & 40                     & Stride = (1, 1), Activation = Linear \\
BatchNorm        & -                    & -                      & -                                    \\
Activation       & -                    & -                      & Activation = Square                  \\
AveragePooling2D & (1, 35)              & -                      & Stride = (1, 7)                      \\
Activation       & -                    & -                      & Activation = log                     \\
Dropout          & -                    & -                      & Rate = 0.5                           \\
Flatten          & -                    & -                      & -                                    \\
Dense            & -                    & -                      & Neurons = 26, Activation = Softmax  \\\hline
\end{tabular}
}
\end{table}

\subsubsection{EEGNet}

It is a compact architecture based on CNN that has been applied for various different BCI tasks. Being a compact architecture, it has the benefit that it can be trained with limited data. Additionally, the architecture produces features that can be neurophysiologically interpreted. The EEGNet architecture can be divided into two stages. In the first stage, $8$ 2D convolutional filers of size $(1,64)$ are applied to the input EEG signals. Subsequently, a depthwise convolution operation of size $(31,1)$ is performed to learn a spatial filter. This two-step process draws inspiration from the filter-bank common spatial pattern (FBCSP) algorithm, which is widely used in EEG applications. Batch normalization is applied after both convolution layers. For the 2D CNN layer, linear activation function is used, while Exponential Linear Unit (ELU) is utilized for the depthwise convolution layer. To regularize the model, dropout with a probability of $0.5$ is further applied. The weights of the spatial filter are also regularized by using a maximum norm constraint of 1 on the weights. An average pooling layer of $(1,4)$ size is applied to the output of this stage.

In the second stage of the architecture, a separable convolution of size $(1,16)$ is applied along with batch normalization and ELU activation function. Another average pooling layer of $(1,8)$ size is applied and the output is flattened to obtain the final feature vector. This is then fed to the output layer comprising $26$ neurons and activated by the softmax activation function. 
The major advantage of EEGNet comes from the use of depthwise and separable convolution layers, which leads to a huge reduction in the parameter count, thereby making the model compact.

\subsubsection{DeepConvNet}

The DeepConvNet was proposed as a generic CNN-based architecture designed with little expert  knowledge to achieve competitive accuracies for EEG-based BCI tasks. The model comprises of $4$ convolution-maxpooling blocks in total. The first block consists of two 2D convolution layers with kernel size $(1,5)$ and $(31,1)$, respectively. This is followed by batch normalization and application of the ELU activation function. Subsequently, maxpooling with kernel size and stride of (1,2) is applied. All the other blocks are identical and comprise of a 2D convolution layer with kernel size $(1,5)$, batch normalization, ELU activation, and maxpooling with kernel size and stride of $(1,2)$. The convolution layers in the first block have $25$ filters, while the number of filters is doubled in each subsequent block. The output of the last block is flattened and fed to the 26-neuron output layer with the softmax activation function. 

\subsubsection{ShallowConvNet}

The ShallowConvNet is a lightweight CNN-based model which also draws inspiration from the FBCSP pipeline. The first two layers of this model can be interpreted as performing a temporal convolution and a spatial filter, respectively. This is analogous to the FBCSP stages of a bandpass filter and Common Spatial Filter. The two 2D convolution layers comprise of $40$ kernels of size $(1,13)$ and $(31,1)$. The larger kernel sizes compared to DeepConvNet allow a larger range of transformations in the convolution layers. The output of the convolution layers is batch normalized, and the square activation function is applied. This is followed by average pooling with kernel size and stride of $(1, 35)$ and $(1, 7)$, respectively, and the application of the logarithmic activation function. These steps are analogous to the trial log-variance computation as done in the FBCSP pipeline. The output is then flattened and fed to the output layer, which is made up of 26 neurons. A dropout with a probability of 0.5 is used to avoid overfitting, while the softmax activation function is used in the output layer. 

\begin{table}[t!]
\caption{Mean recognition accuracies for the EEG-based airwriting recognition task by using Processed EEG, ICA Components, and Scout time series. The effect on accuracies by using different frequency bands in conjunction with different Deep Learning models is also depicted.}
\label{tab:feats}
\centering
\scalebox{0.64}{
\begin{tabular}{ccccccccc}
\hline
\multicolumn{1}{l}{}                 & \multicolumn{1}{l}{} & \multicolumn{7}{c}{\textbf{Frequency Band}}                                                                                  \\\hline\hline
                                     & \textbf{Model}       & \textbf{Delta} & \textbf{Theta} & \textbf{Alpha} & \textbf{Beta} & \textbf{Gamma} & \textbf{Combined} & \textbf{Delta+Theta} \\\hline
{\textbf{EEG (V)}}    & \textbf{EEGNet}      & 37.35          & 32.93          & 24.24          & 17.18         & 10.37          & 39.37             & 40.05                \\
                                     & \textbf{DeepConvNet}       & 32.25          & 27.32          & 20.06          & 12.51         & 7.87           & 37.22             & 37.80                \\
                                     & \textbf{ShallowConvNet}       & 17.34          & 20.04          & 13.50          & 12.81         & 9.47           & 23.88             & 22.72                \\\hline\hline
{\textbf{ICA (U)}}    & \textbf{EEGNet}      & 41.77          & 35.75          & 24.89          & 17.28         & 11.94          & 43.85             & 44.04                \\
                                     & \textbf{DeepConvNet}       & 36.93          & 29.96          & 20.80          & 13.07         & 7.89           & 42.56             & 42.52                \\
                                     & \textbf{ShallowConvNet}       & 23.50          & 23.45          & 15.07          & 13.76         & 9.97           & 27.97             & 26.36                \\\hline\hline
{\textbf{Source (S)}} & \textbf{EEGNet}      & 37.34          & 33.11          & 24.40          & 16.53         & 10.28          & 39.40             & 40.23                \\
                                     & \textbf{DeepConvNet}       & 32.37          & 27.40          & 20.10          & 12.22         & 7.42           & 38.57             & 38.19                \\
                                     & \textbf{ShallowConvNet}       & 16.49          & 19.36          & 12.96          & 12.70         & 9.47           & 21.51             & 20.70   \\\hline            
\end{tabular}
}
\end{table}

\begin{table}[t!]
\caption{Mean recognition accuracies for the EEG-based airwriting recognition task by using Spherical Harmonic Decomposition based features $[V^{SH}_{nm}]$. The effect on accuracies by using different decomposition order and frequency bands in conjunction with different Deep Learning models is also depicted.}
\label{tab:spherical}
\centering
\scalebox{0.68}{
\begin{tabular}{ccccccccc}
\hline
                                   &                 & \multicolumn{7}{c}{\textbf{Frequency Band}}                                                                                  \\\hline\hline
                                   & \textbf{Model}  & \textbf{Delta} & \textbf{Theta} & \textbf{Alpha} & \textbf{Beta} & \textbf{Gamma} & \textbf{Combined} & \textbf{Delta+Theta} \\\hline
{\textbf{$[V^{SH}_{nm}]$ (N=4)}} & \textbf{EEGNet} & 36.53          & 32.04          & 23.44          & 16.34         & 10.27          & 38.85             & 39.68                \\
                                   & \textbf{DCNet}  & 31.54          & 27.17          & 20.00          & 11.96         & 7.77           & 36.94             & 37.18                \\
                                   & \textbf{SCNet}  & 17.21          & 19.07          & 13.42          & 12.57         & 9.22           & 22.52             & 21.90                \\\hline\hline
{\textbf{$[V^{SH}_{nm}]$ (N=3)}} & \textbf{EEGNet} & 33.20          & 28.63          & 21.20          & 14.89         & 9.87           & 36.36             & 37.00                \\
                                   & \textbf{DCNet}  & 28.93          & 24.08          & 17.65          & 11.15         & 7.81           & 34.11             & 34.19                \\
                                   & \textbf{SCNet}  & 15.69          & 16.44          & 12.08          & 11.50         & 8.61           & 20.07             & 19.41                \\\hline\hline
{\textbf{$[V^{SH}_{nm}]$ (N=2)}} & \textbf{EEGNet} & 29.13          & 24.97          & 19.39          & 13.08         & 9.06           & 33.16             & 32.70                \\
                                   & \textbf{DCNet}  & 25.43          & 21.32          & 16.46          & 10.42         & 7.43           & 30.68             & 30.21                \\
                                   & \textbf{SCNet}  & 14.15          & 13.84          & 10.63          & 10.88         & 8.61           & 17.49             & 17.02            \\\hline   
\end{tabular}
}
\end{table}
 
\section{Experimental Setup and Results}

\subsection{Experimental Details}

The processed signals i.e. $[V], [U], [S], [V^{SH}_{nm}],$ and $[V^{H^2}_{nm}]$ were used as input to the deep learning models for airwiriting recognition. Segments of the signals
were taken from the time windows $-1$ to $2$ seconds as inspired from \cite{pei2021online}. For letters that are written in a duration of less than $2$ seconds, zero padding was done to fix the length of the input. Since the EEG signals were recorded at a sampling rate of $500$ Hz, the length of the input feature matrix was $1500$. The input, therefore, was a 2D matrix of dimension $\kappa \times 1500$, where $\kappa$ depends on the feature being used ($31$ for $[V]$ and $[U]$, $62$ for $[S]$, $(N+1)^2$ for $[V^{SH}_{nm}],$ and $[V^{H^2}_{nm}]$). The order $N$ was varied from 2 to 4, leading to reduced dimensionality ($\kappa\in\{9,16,25\}$) in the transformed domain. Subsequently, the input matrix is normalized by using the standard z-normalization technique to make the input data have zero mean and unit variance. 

In order to validate the efficacy of the proposed frameworks, a user dependent $10$-fold cross-validation strategy is adopted. For each subject, the $100$ trials per character are split into $90$ trials for training the deep learning model and $10$ trials for testing the accuracy of the trained deep learning model. This process is repeated $10$ times to ensure that each trial is a part of the test set once. Additionally, for each fold, $20\%$ of the training data is separately used as the validation set.  The accuracies of each of the $10$ folds are averaged to give the EEG-based airwriting recognition accuracy for the subject. The average of accuracy scores of all the $10$ subjects is then computed and reported. The deep learning models are trained by employing a mini-batch training process with a batch size of $128$. The cross-entropy loss is minimized using the Adam optimizer in order to learn the model parameters. The model is trained for a maximum of $500$ epochs, and early stopping with patience of $20$ conditioned on validation set accuracy is also employed to prevent overfitting. 

\begin{table}[t!]
\caption{Mean recognition accuracies for the EEG-based airwriting recognition task by using Head Harmonic Decomposition based features $[V^{H^2}_{nm}]$. The effect on accuracies by using different decomposition order and frequency bands in conjunction with different Deep Learning models is also depicted.}
\label{tab:head}
\centering
\scalebox{0.68}{
\begin{tabular}{ccccccccc}
\hline
                                   &                 & \multicolumn{7}{c}{\textbf{Frequency Band}}                                                                                  \\\hline\hline
                                   & \textbf{Model}  & \textbf{Delta} & \textbf{Theta} & \textbf{Alpha} & \textbf{Beta} & \textbf{Gamma} & \textbf{Combined} & \textbf{Delta+Theta} \\\hline
{\textbf{$[V^{H^2}_{nm}]$ (N=4)}} & \textbf{EEGNet} & 36.82          & 33.03          & 23.29          & 16.37         & 10.21          & 39.43             & 40.02                \\
                                   & \textbf{DCNet}  & 31.59          & 27.07          & 19.94          & 12.10         & 7.66           & 37.42             & 37.41                \\
                                   & \textbf{SCNet}  & 17.33          & 19.67          & 13.27          & 12.62         & 8.90           & 22.47             & 22.07                \\\hline\hline
{\textbf{$[V^{H^2}_{nm}]$ (N=3)}} & \textbf{EEGNet} & 34.85          & 29.38          & 22.32          & 14.88         & 10.10          & 37.69             & 37.52                \\
                                   & \textbf{DCNet}  & 30.54          & 25.47          & 18.56          & 11.90         & 7.28           & 35.47             & 35.72                \\
                                   & \textbf{SCNet}  & 16.53          & 17.83          & 13.02          & 12.20         & 8.96           & 21.24             & 21.17                \\\hline\hline
{\textbf{$[V^{H^2}_{nm}]$ (N=2)}} & \textbf{EEGNet} & 29.77          & 25.15          & 19.87          & 13.67         & 9.55           & 33.31             & 33.91                \\
                                   & \textbf{DCNet}  & 25.68          & 21.57          & 16.42          & 10.72         & 7.13           & 31.25             & 30.90                \\
                                   & \textbf{SCNet}  & 14.76          & 14.70          & 11.23          & 10.99         & 8.45           & 18.37             & 18.03        \\\hline       
\end{tabular}
}
\end{table}

\begin{figure}[!t]
    \centering{
    \includegraphics[width=\linewidth]{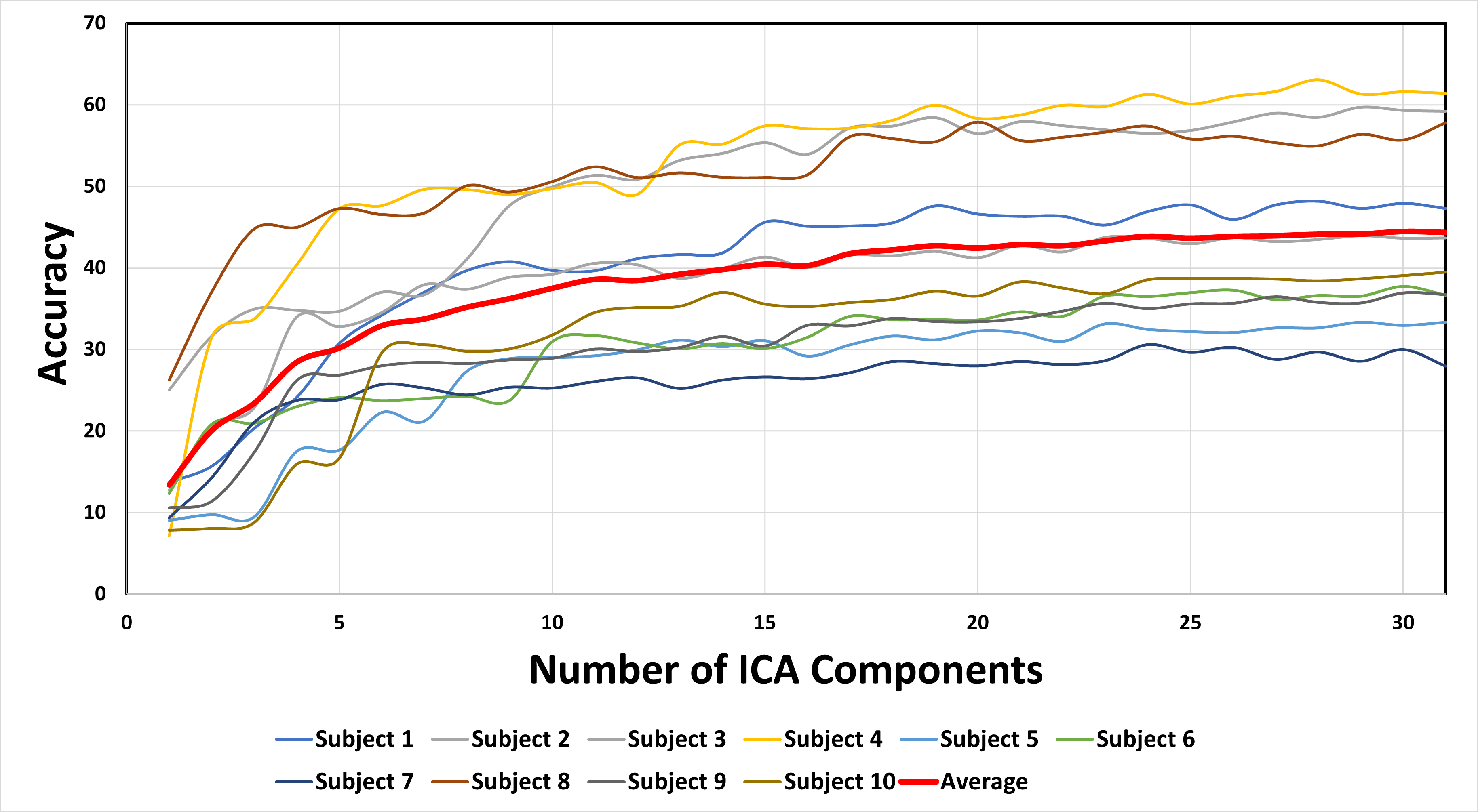}}
    \caption{Effect of variation of the number of ICA Components on the mean recognition accuracies for the EEG-based airwriting recognition system using the EEGNet model. The accuracies for individual subjects and the mean accuracy are depicted.}
    \label{fig:ICACompVariation}
\end{figure}

\subsection{Results}

In the first set of experiments, the performance of the EEG-based airwriting recognition system by using different feature inputs: processed EEG ([$V$]), ICA Components ([$U$]), and source-domain scout time series ([$S$]) is explored. The mean recognition accuracies for the $10$ subjects by using the three different model architectures are presented in Table \ref{tab:feats}. Additionally, the effect of different frequency bands on the performance of the airwriting recognition system is also presented in Table \ref{tab:feats}. Subsequently, analysis is performed to identify whether a reduced set of ICA components can be utilized for effectively categorizing airwritten alphabets. For this, the ICA components are sorted in decreasing order of variance and the recognition accuracy is obtained by using the first $k \in [1,31]$ components. For this purpose, ICA components filtered in the Delta$+$Theta frequency range are considered in conjunction with the EEGNet model for classification. The variation of accuracy scores for each of the $10$ subjects and the mean accuracy is presented in Figure \ref{fig:ICACompVariation}. 

Further, the performance of spherical harmonic and head harmonic decomposition-based features in identifying airwritten characters is also explored. The effect of the different frequency bands and deep learning models is comprehensively explored for this purpose. Since the order of harmonic decomposition is a variable constrained by the equation $N\leq(\sqrt{I}-1)$, the effect of varying $N\in[2,3,4]$ is also explored. The mean recognition accuracies for spherical harmonic and head harmonic decomposition-based features are presented in Tables \ref{tab:spherical} and \ref{tab:head}, respectively.

\subsection{Discussion}

\subsubsection{Deep Learning Model Analysis}

In this work, three different deep learning architectures, namely EEGNet, DeepConvNet, and ShallowConvNet, were utilized to assess the performance of an EEG-based airwriting recognition system. As evident from the results in \cref{tab:feats,tab:spherical,tab:head}, EEGNet architecture outperforms the other models, suggesting its suitability for airwriting recognition. This implies that the EEGNet architecture is most suited to be utilized for the task of airwriting recognition. The superior performance of EEGNet can be attributed to the limited amount of training data available. This architecture incorporates depthwise and separable convolutions, which are specifically designed for EEG data and integrates the feature extraction process within end-to-end training. Additionally, EEGNet has the fewest parameters among the models used in this study, making it a lightweight and robust choice for learning interpretable features for the task of airwriting recognition.  Subsequently, a detailed one-tailed t-test was conducted to compare the performance of the different models when processed EEG data was used as input. The results of the t-test, presented in Table \ref{tab:ttestmodels}, provide further insights into the statistical significance of the performance differences between the models. 

The utilization of classical machine learning algorithms in combination with handcrafted features resulted in inadequate performance when attempting to identify airwritten alphabets from EEG signals. For instance, an SVM model trained using a set of handcrafted features per EEG channel achieved a meager accuracy of $4.12\%$. Similar trends were observed with other classical algorithms like logistic regression and decision trees. The poor performance of these classical ML algorithms can be attributed to their failure to capture the intricate dynamics inherent in EEG signals, which are crucial for accurately recognizing airwritten alphabets. These algorithms lack the ability to effectively extract the complex and nuanced information present in the EEG data, leading to suboptimal results. Therefore, alternative approaches that can capture the intricate dynamics of EEG signals, such as deep learning models, are necessary to improve the performance of airwriting recognition systems.

\subsubsection{EEG feature analysis}

To assess the effectiveness of the proposed airwriting recognition system and determine the most suitable features for the task, processed EEG, ICA components, and source-domain scout time series were employed as inputs to the deep learning models. The results in Table \ref{tab:feats} indicate that the highest recognition accuracies were achieved when using ICA components as input. These components represent independent sources, as the ICA algorithm acts as a spatial filter. Consequently, the individual ICA components serve as spatially filtered versions of the EEG signal, leading to improved recognition accuracies. This observation aligns with a similar trend reported in \cite{pei2021online}, where ICA components yielded the best accuracy in a handwriting recognition task. Moreover, Table \ref{tab:feats} reveals that the performance of the airwriting recognition system is comparable when using source-domain scout time series instead of processed EEG signals. This suggests that the scout time-series features capture relevant information for the task. The results of a detailed one-tailed t-test, presented in Table \ref{tab:ttestfeats}, show a statistically significant difference between ICA components and processed EEG or scout time-series features. However, no significant differences were observed among other feature pairs. In summary, the utilization of ICA components as input to the deep learning models resulted in the highest recognition accuracies in the airwriting recognition system. Additionally, the scout time-series features demonstrated similar performance to processed EEG signals. The statistical analysis confirmed the significance of the difference between ICA components and the other features, highlighting the effectiveness of ICA for this task.

\begin{table}[t!]
\caption{p-values of the comparative one-tailed t-test between different deep learning models. Entries with a significance value below 0.05 are depicted in bold.}
\label{tab:ttestmodels}
\centering
\scalebox{1.0}{
\begin{tabular}{|c|c|c|c|}
\hline 
               & EEGNet            & DeepConvNet       & ShallowConvNet    \\\hline 
EEGNet         & -                 & \textbf{0.0009}   & \textbf{1.08$\times 10^{-7}$} \\\hline 
DeepConvNet    & \textbf{0.0009}   & -                 & \textbf{1.43$\times 10^{-7}$} \\\hline 
ShallowConvNet & \textbf{1.08$\times 10^{-7}$} & \textbf{1.43$\times 10^{-7}$} & -  \\\hline              
\end{tabular}
}
\end{table}

\begin{table}[t!]
\caption{p-values of the comparative one-tailed t-test between different features used for EEG-based airwriting recognition using the EEGNet model. Entries with a significance value below 0.05 are depicted in bold.}
\label{tab:ttestfeats}
\centering
\scalebox{1.0}{
\begin{tabular}{|c|c|c|c|c|c|}
\hline
       & $[V]$             & $[U]$             & $[S]$        & $[V^{SH}_{nm}]$              & $[V^{H^2}_{nm}]$              \\\hline
$[V]$     & -               & \textbf{0.0011} & 0.4636          & 0.1853          & 0.4296          \\\hline
$[U]$    & \textbf{0.0011} & -               & \textbf{0.0003} & \textbf{0.0007} & \textbf{0.0013} \\\hline
$[S]$ & 0.4636          & \textbf{0.0003} & -               & 0.1613          & 0.4666          \\\hline
 $[V^{SH}_{nm}]$     & 0.1853          & \textbf{0.0007} & 0.1613          & -               & 0.0838          \\\hline
$[V^{H^2}_{nm}]$    & 0.4296          & \textbf{0.0013} & 0.4666          & 0.0838          & -  \\\hline            
\end{tabular}
}
\end{table}

\begin{table}[t!]
\caption{p-values of the comparative one-tailed t-test between different frequency bands used for EEG-based airwriting recognition using the EEGNet model and ICA source activity matrix. Entries with a significance value below 0.05 are depicted in bold.}
\label{tab:ttestfreq}
\centering
\scalebox{0.6}{
\begin{tabular}{|c|c|c|c|c|c|c|c|}
\hline
            & Delta             & Theta             & Alpha             & Beta              & Gamma             & Complete          & Delta+Theta       \\\hline
Delta       & -                 & \textbf{0.0096}   & \textbf{0.0003}   & \textbf{1.31$\times 10^{-5}$} & \textbf{5.7$\times 10^{-6}$}  & 0.0111            & \textbf{1.6$\times 10^{-5}$}  \\\hline
Theta       & \textbf{0.0096}   & -                 & \textbf{0.0006}   & \textbf{2.38$\times 10^{-5}$} & \textbf{1.15$\times 10^{-5}$} & \textbf{0.0018}   & \textbf{0.0016}   \\\hline
Alpha       & \textbf{0.0003}   & \textbf{0.0006}   & -                 & \textbf{5.3$\times 10^{-7}$}  & \textbf{7.71$\times 10^{-7}$} & \textbf{3.74$\times 10^{-5}$} & \textbf{6.37$\times 10^{-5}$} \\\hline
Beta        & \textbf{1.31$\times 10^{-5}$} & \textbf{2.38$\times 10^{-5}$} & \textbf{5.3$\times 10^{-7}$}  & -                 & \textbf{0.0004}   & \textbf{2.24$\times 10^{-6}$} & \textbf{3.48$\times 10^{-6}$} \\\hline
Gamma       & \textbf{5.7$\times 10^{-6}$}  & \textbf{1.15$\times 10^{-5}$} & \textbf{7.71$\times 10^{-7}$} & \textbf{0.0004}   & -                 & \textbf{1.08$\times 10^{-6}$} & \textbf{1.76$\times 10^{-6}$} \\\hline
Complete    & 0.0111            & \textbf{0.0018}   & \textbf{3.74$\times 10^{-5}$} & \textbf{2.24$\times 10^{-6}$} & \textbf{1.08$\times 10^{-6}$} & -                 & 0.3916            \\\hline
Delta+Theta & \textbf{1.6$\times 10^{-5}$}  & \textbf{0.0016}   & \textbf{6.37$\times 10^{-5}$} & \textbf{3.48$\times 10^{-6}$} & \textbf{1.76$\times 10^{-6}$} & 0.3916            & -      \\\hline          
\end{tabular}
}
\end{table}

\subsubsection{Harmonic Decomposition Analysis}

The spherical and head harmonic decomposition techniques were employed to break down the EEG signals into functions defined over the surface of a sphere or the head. These decomposed features were then used as input for the deep learning models in the airwriting recognition task. The results in \cref{tab:spherical,tab:head} demonstrate that the recognition accuracies achieved using this approach are comparable to those obtained by using processed EEG as input. This finding is also supported by the t-test results presented in Table \ref{tab:ttestfeats}. One advantage of utilizing the decomposed features is the reduction in dimensionality ($31$ to $\kappa \in \{9,16,25\}$) compared to using the original EEG signals as input. This reduction can potentially simplify the learning process and improve the efficiency of the deep learning models. However, it should be noted that decreasing the order of decomposition, represented by the parameter $N \in [2,3,4]$, leads to a decrease in the mean recognition accuracy. This suggests that a higher order of decomposition captures more relevant information for accurate airwriting recognition. Overall, the use of spherical and head harmonic decomposition techniques provides a viable alternative for representing EEG signals in the airwriting recognition task. These features offer dimensionality reduction benefits while maintaining comparable recognition accuracies to those achieved with processed EEG input.

\subsubsection{Number of ICA Components analysis}

Figure \ref{fig:ICACompVariation} depicts the relationship between the recognition accuracy and the number of ICA components used as input to the deep learning model. The components were sorted in descending order of variance for this analysis. The results indicate that initially, as the number of components increases, the accuracy of the airwriting recognition system improves. However, this trend eventually levels off, and the mean recognition accuracy plateaus at around 20 components. This observation suggests that a lower number of components can be employed without significantly compromising the recognition accuracy. This analysis provides valuable insights into reducing the dimensionality of the input to the deep learning model. By utilizing a smaller number of components, the overall number of model parameters can be reduced, resulting in a lightweight model. Consequently, this approach offers the potential to maintain a high recognition accuracy while optimizing computational efficiency.

\subsubsection{EEG frequency band analysis}

In this study, an analysis of EEG frequency bands was conducted to determine the impact of different frequency ranges on the performance of the airwriting recognition system. The results presented in \cref{tab:feats,tab:spherical,tab:head} indicate that the features filtered in the low-frequency bands achieve the highest accuracy. Specifically, the Delta frequency band exhibits the highest accuracy among individual frequency bands, followed by the Theta band. Also, it is noted that using the signals filtered in the Delta$+$Theta frequency bands yields the best accuracy.   Conversely, when higher frequency bands are utilized, the system's performance declines. The significance values obtained from the pairwise one-tailed t-test, as shown in Table \ref{tab:ttestfreq}, further support the observation that the low-frequency bands are associated with superior accuracy. This finding is consistent with previous studies such as \cite{pei2021online, kobler2018tuning}, which have also demonstrated the dominance of low-frequency bands in capturing movement dynamics. Thus, the results obtained in this study align with the existing body of research and confirm the efficacy of utilizing low-frequency bands for the airwriting recognition task.

\section{Conclusion}

In this paper, the effectiveness of a non-invasive brain-computer interface for recognizing airwritten characters was explored. First, the NeuroAiR dataset comprising $31$-channel scalp-recorded EEG signals taken from $10$ subjects while writing $100$ repetitions of uppercase English alphabets was constructed. To the best of the author's knowledge, this is the first attempt at creating such a dataset for the EEG-based airwriting recognition task. Several different features, such as processed EEG, Independent Component Analysis (ICA) components, and source domain-based scout time series, were used in conjunction with three different deep learning models for the $26$-class airwriting classification task. Furthermore, spherical harmonic and head harmonic-based decomposition schemes were adopted to extract features which were then utilized for the airwriting recognition task. The effect of different EEG-frequency bands: Delta, Theta, Alpha, Beta, Gamma, Delta$+$Theta, and the combined spectrum information on the recognition accuracy was also explored. Among the different features, the ICA components, when used  as input to the EEGNet classification model, yielded the best accuracy of $44.04\%$ when the data was filtered in the Delta$+$Theta band. Analysis of different frequency bands revealed that the low-frequency ranges yield improved performance, which is consistent with the literature, as low-frequency bands are associated with movement dynamics. Future work in this area may be focused on further enhancing system performance through the exploration of sophisticated feature extraction methods and advanced deep learning techniques. In conclusion, the reported recognition accuracies for the EEG-based airwriting task in this study serve as a strong baseline for future research endeavors in this domain. The findings have the potential to establish airwriting as an alternative input method for applications in Human-Computer Interaction, opening up new avenues for seamless interaction between individuals and technology.

\section*{Acknowledgment}

The authors would like to thank Prof. Saurabh Gandhi, Shivani Ranjan, and Anant Jain for their valuable insights during the data collection setup and all the subjects for their participation in the experiment.

\bibliographystyle{IEEEtran}
\bibliography{refs.bib}

\end{document}